\documentclass[stu,floatsintext,donotrepeattitle]{apa7}
\usepackage{url}
\usepackage{graphicx}
\usepackage{array}  
\usepackage{subcaption} 
\usepackage{amsmath} 
\usepackage{ragged2e}
\usepackage{tabularx}
\usepackage{pgffor}    
\usepackage{float}

\newenvironment{indented}{\setlength{\parindent}{0.5in}}{}

\makeatletter
\renewcommand\paragraph{\@startsection{paragraph}{4}{\z@}{1.5ex plus .5ex minus .2ex}{1em}{\normalfont\normalsize\bfseries}}
\makeatother

\title{\textbf{Pop-out vs. Glue: A Study on the pre-attentive and focused attention stages in Visual Search tasks}}
\authorsnames{Hendrik Beukelman, Wilder C. Rodrigues}
\authorsaffiliations{Faculty of Humanities, Utrecht University}
\course{Artificial Intelligence}
\professor{Krista Overvliet, Baptist Liefooghe}
\duedate{\today}

\begin{document}
\maketitle
\justifying

\section*{\textbf{Abstract}}
This study explores visual search asymmetry and the detection process between parallel and serial search strategies, building upon Treisman's Feature Integration Theory \cite{treisman1}. Our experiment examines how easy it is to locate an oblique line among vertical distractors versus a vertical line among oblique distractors, a framework previously validated by Treisman \& Gormican (1988) \cite{treisman2} and Gupta et al. (2015) \cite{gupta}. We hypothesised that an oblique target among vertical lines would produce a perceptual 'pop-out' effect, allowing for faster, parallel search, while the reverse condition would require serial search strategy. Seventy-eight participants from Utrecht University engaged in trials with varied target-distractor orientations and number of items. We measured reaction times and found a significant effect of target type on search speed: oblique targets were identified more quickly, reflecting 'pop-out' behaviour, while vertical targets demanded focused attention ('glue phase'). Our results align with past findings, supporting our hypothesis on search asymmetry and its dependency on distinct visual features. Future research could benefit from eye-tracking and neural network analysis, particularly for identifying the neural processing of visual features in both parallel and serial search conditions.
\clearpage
\section{Introduction}
Humans regularly face visually demanding tasks, such as locating objects or people in crowded settings, which are vital to navigating uncertainties in daily life. Success in these tasks depends on factors like awareness, cognitive abilities, and the nature of the search itself.\\
\begin{indented}
Some studies have explored the complexities of visual search, focusing on asymmetry, where locating target A among distractors B is easier than finding B among A. Our research specifically examines the asymmetry between finding an oblique line among straight lines versus a straight line among oblique lines.\\
Anne Treisman's study (Treisman \& Gelade, 1980) \cite{treisman1} found that certain features, like colour, are more easily detected than others, such as orientation. Further, Treisman \& Gormican (1988) \cite{treisman2} showed that identifying a vertical target among oblique distractors took longer than identifying an oblique target among vertical distractors, this supports the idea that a basic feature enhances detection.\\
We aim to replicate these findings with the following research question:
\textit{Does searching for an oblique target among vertical distractors result in search asymmetry, and vice versa?}\\
We anticipate a 'pop-out' effect when participants search for an oblique target among vertical distractors, suggesting a parallel search. As opposed to a serial search pattern in the reverse condition. Consistent with Treisman \& Gormican’s findings \cite{treisman2}, we predict faster identification of oblique targets, aligning with the 'pop-out' effect, while vertical targets will require focused attention ('glue' phase), particularly as distractor numbers increase.
\end{indented}
\section{Method}
As inspiration, and also for reproducibility purposes, we took the foundation established by other researchers in the fields of Visual Search and Feature Integration (Gupta et al \cite{gupta}, Treisman \& Gelade \cite{treisman1}) to setup our study. In the next sections, we will go over the whole process of understanding participants' selection, their demographic details, and then diving into the stimuli used, along with the design of the experiment and the procedure behind it.
\subsection{Participants}
\begin{figure}[h!]
    \centering
    \caption{}
    \includegraphics[width=0.8\textwidth]{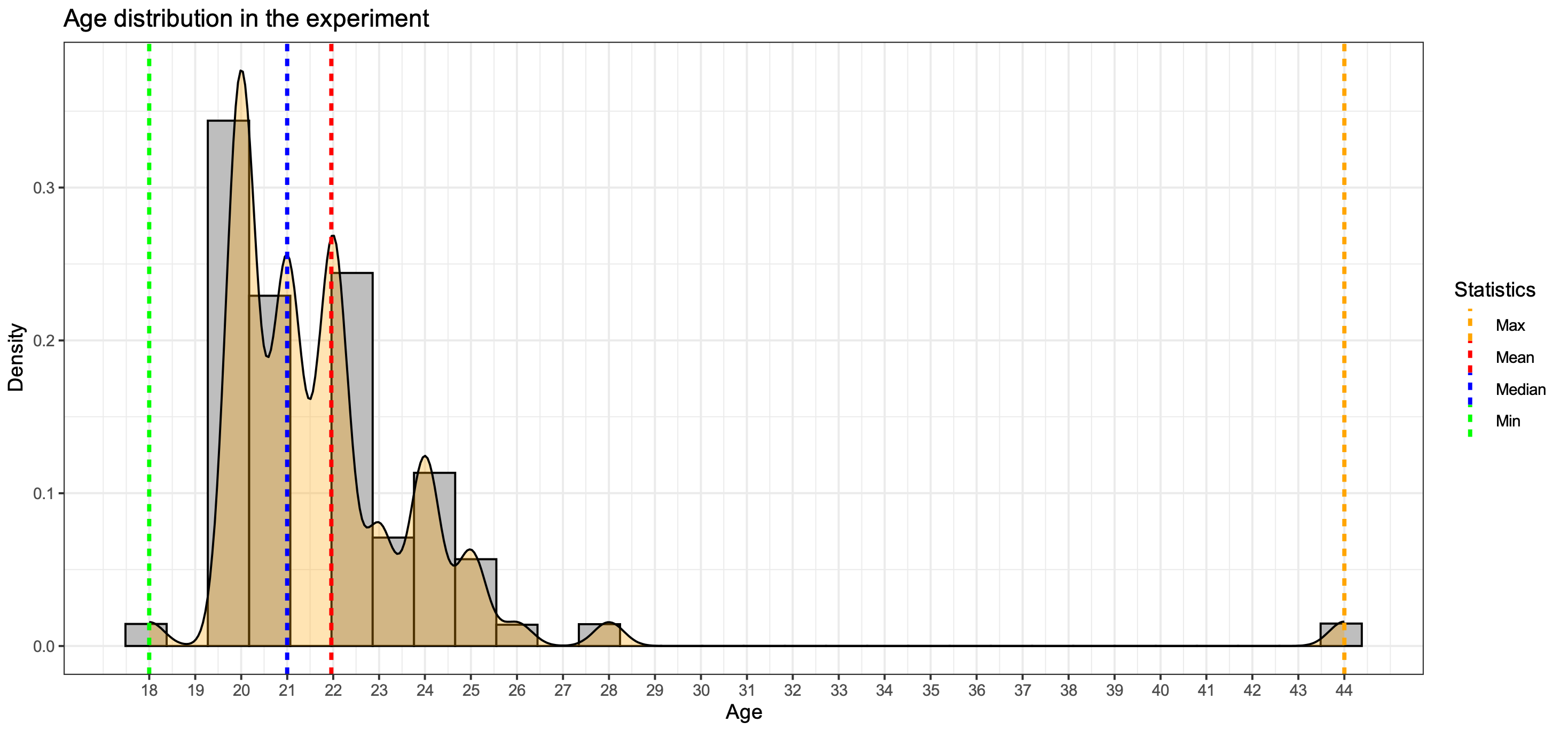}
    \label{fig-age-hist}
\end{figure}

To conduct this study, we gathered data from 78 participants. Those were all students from the Utrecht University, following the Scientific Method and Statistics course from the Faculty of Humanities. Although we gave an option to the participants to inform their gender, most of them opted out. This made any statistical study based on gender irrelevant. From the 78 participants, 5.13\% (4) were left-handed and 94.87\% (74) were right-handed. Extra demographic information is available in Figure \ref{fig-age-hist}.\\
\begin{indented}
The experiment was held at the Science Park campus, located at the the Utrecht University, in the city of Utrecht, the Netherlands. The participants were asked to read and sign, upon agreement, our participation form. The study was conducted in line with ethical principals governed by the Code of Ethics of the World Medical Association (Declaration of Helsinki).
\end{indented}
\subsection{Stimuli \& Devices}
\subsubsection{Stimuli}
In the design of the experiment we followed a similar setup as the one used by Gupta et al \cite{gupta} in their Visual Search Asymmetry publication, experiment 5, that tackled oriented Targets (T) and Distractions (D). With respect to the amount of elements presented to the participants and the number of items per trial, our experiment was also similar to Gupta's. For instance, we used trial blocks with 1, 4, 8, and 12 items. Per each set we had one target, with the remaining items in the set being distractors. The Equation \ref{eq1} depicts what a set \textit{S} was comprised of.
\begin{equation}
S = \{ N \mid N = nD + 1T \}
\label{eq1}
\end{equation}

\begin{indented}
We setup our experiment with two conditions where the distractors and target items would be swapped. During the experiment, the participant is presented with a first, randomly selected condition, that will have either an oblique line with a fixed angle as a target and several (up to 11) vertical lines that are distractors, and vice-versa, as depicted in Figure \ref{fig:1}.
\begin{figure}[h!]
    \centering
    \caption{Visual search task with 2 conditions: search for oblique lines with fixed angles among vertical lines and vice versa.}
    \begin{subfigure}[b]{0.45\textwidth}
        \centering
        \includegraphics[width=\textwidth]{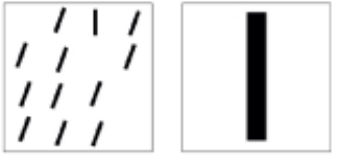}
        \caption{Target (vertical line) among distractors, oblique with a fixed angle lines.}
        \label{fig:1a}
    \end{subfigure}
    \hfill
    \begin{subfigure}[b]{0.45\textwidth}
        \centering
        \includegraphics[width=\textwidth]{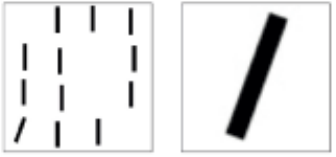}
        \caption{Target (oblique line) among distractors, vertical lines.}
        \label{fig:1b}
    \end{subfigure}
    \begin{subfigure}[b]{0.45\textwidth}
        \centering
        \includegraphics[width=\textwidth]{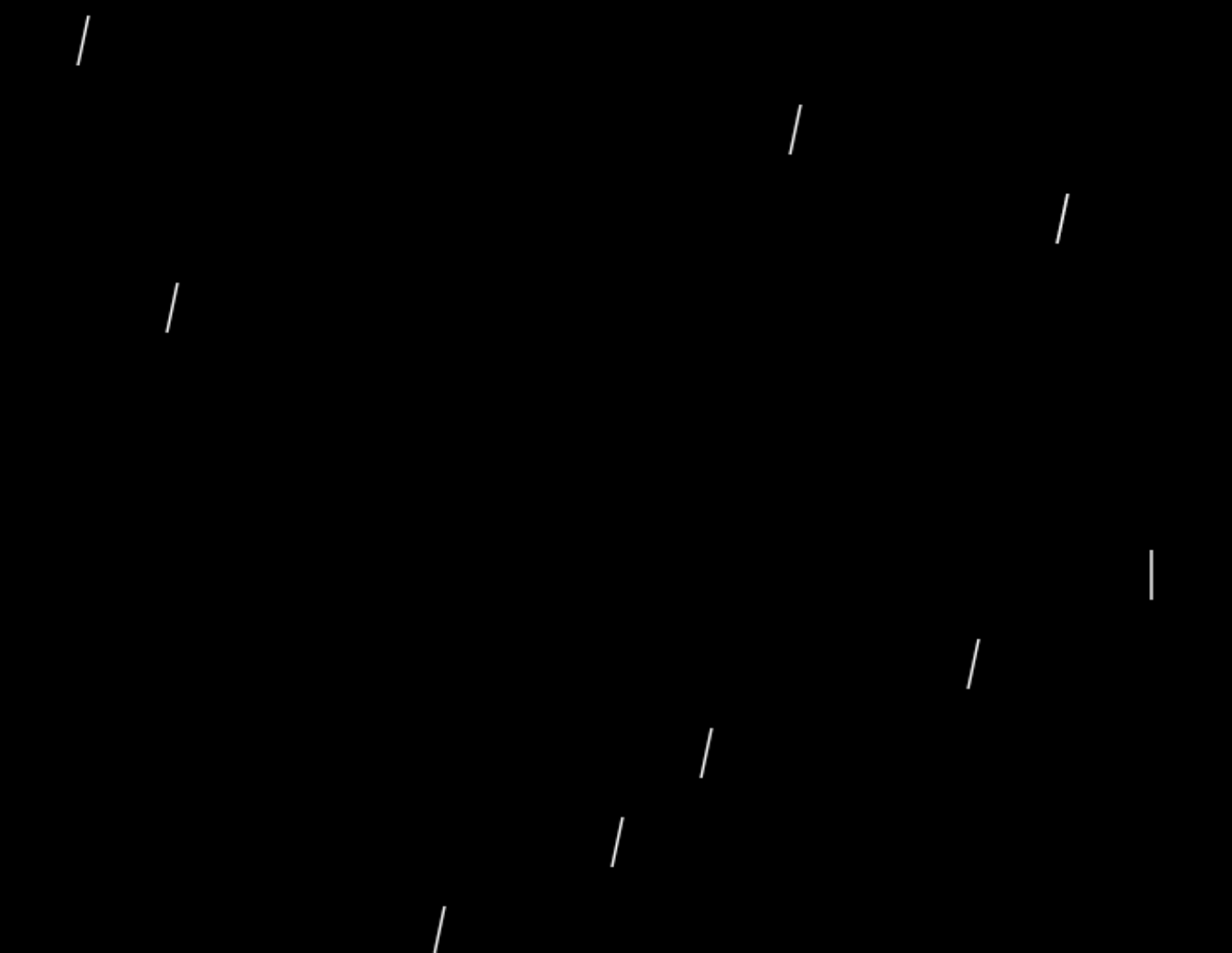}
        \caption{Target (vertical line) among distractors, oblique with a fixed angle lines used during our experiment.}
        \label{fig:1c}
    \end{subfigure}
    \label{fig:1}
\end{figure}

Our version of the experiment differs from Gupta's in the following terms: 1) We do not limit the area where the target and distractors are placed to a 4 x 4 grid. Instead, we display the elements randomly using the whole of a 14" inch screen; 2) We use a black background with white elements. Those differences are depicted in Figure \ref{fig:1c}.\\
\end{indented}
\subsubsection{Devices}
We developed the experiment using PsychoPy, a tool used for stimulus generation and experimental control. PsychoP is published by Open Science Tools Ltd.\\
\begin{indented}
We used an Apple MacBook (see Table \ref{apple-tb}) for the development of the experiment. The experiment was executed on HP Elite 840 G3 laptops at the campus. Those laptops had a 14-inch screen with a resolution of 1920x1080 pixels with a refresh rate was set 60 Hertz.
\begin{table}[h!]
\centering
\caption{Technical specifications of the computer used for the development of the experiment.}
\begin{tabularx}{\textwidth}{|>{\centering\arraybackslash}X|>{\centering\arraybackslash}X|>{\centering\arraybackslash}X|>{\centering\arraybackslash}X|>{\centering\arraybackslash}X|>{\centering\arraybackslash}X|}
\hline
\textbf{Brand} & \textbf{Processor} & \textbf{CPU Cores} & \textbf{Memory} & \textbf{Operating System} & \textbf{Display}\\
\hline
Apple & Apple Silicon M3 & 11 & 36 GB & macOS v15.0 & Retina XDR 14-inch \\ \hline
\end{tabularx}
\label{apple-tb}
\end{table}
\end{indented}
\subsection{Experiment design \& procedure}
Each of the trials started with a fixation cross, used to ensure all participants have the same starting point. The participants must click on the cross to proceed. After that, a blank screen was displayed for the duration of 200 milliseconds. The participants would then be prompted with the first block of trials displaying 1, 4, 8 and 12 items per trial. The selection of the which target and distractors to use was applied randomly. So, the first block could either be vertical targets with oblique distractors or vice-versa.

\begin{indented}
The Reaction Time (RT) of the participants was measured from the moment the screen, containing target and distractors, was presented until they took action and clicked on the the target.\\
The setup of the experiment is depicted in Figure \ref{fig-psy}, where the full loop, from the Welcome screen to the final Thank you message, is seen. There are 2 blocks: 1) for the trials repetition (e.g. 1, 4, 8, and 12 items per trial); and 2) for the second condition. For example, if the experiment starts with oblique target and vertical distractors, then the second condition will be the other way around. This selection is done randomly. This approach was adopted to avoid confound issues.\\
\begin{figure}[h!]
    \centering
    \caption{Experiment setup depicted as a PsychoPy flow diagram, containing 2 outer loops for different trials setup.}
    \includegraphics[width=0.8\textwidth]{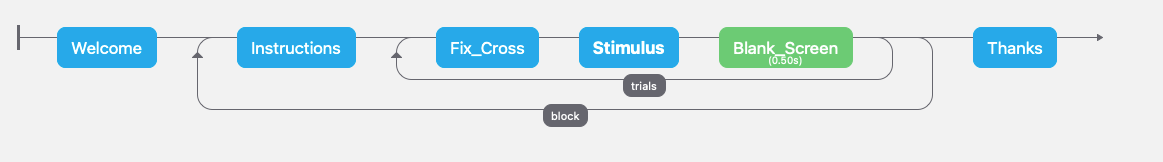}
    \label{fig-psy}
\end{figure}
In Figure \ref{fig-exp-sketch} we show a sketch of how the experiment has been presented to the participants. We omitted the \textit{Welcome}, \textit{Instruction} and \textit{Thanks} routines from the sketch for the sake of simplicity.
\begin{figure}[h!]
    \centering
    \caption{Sketch depicting the trial blocks for the 2 conditions used in the experiment.}
    \includegraphics[width=0.8\textwidth]{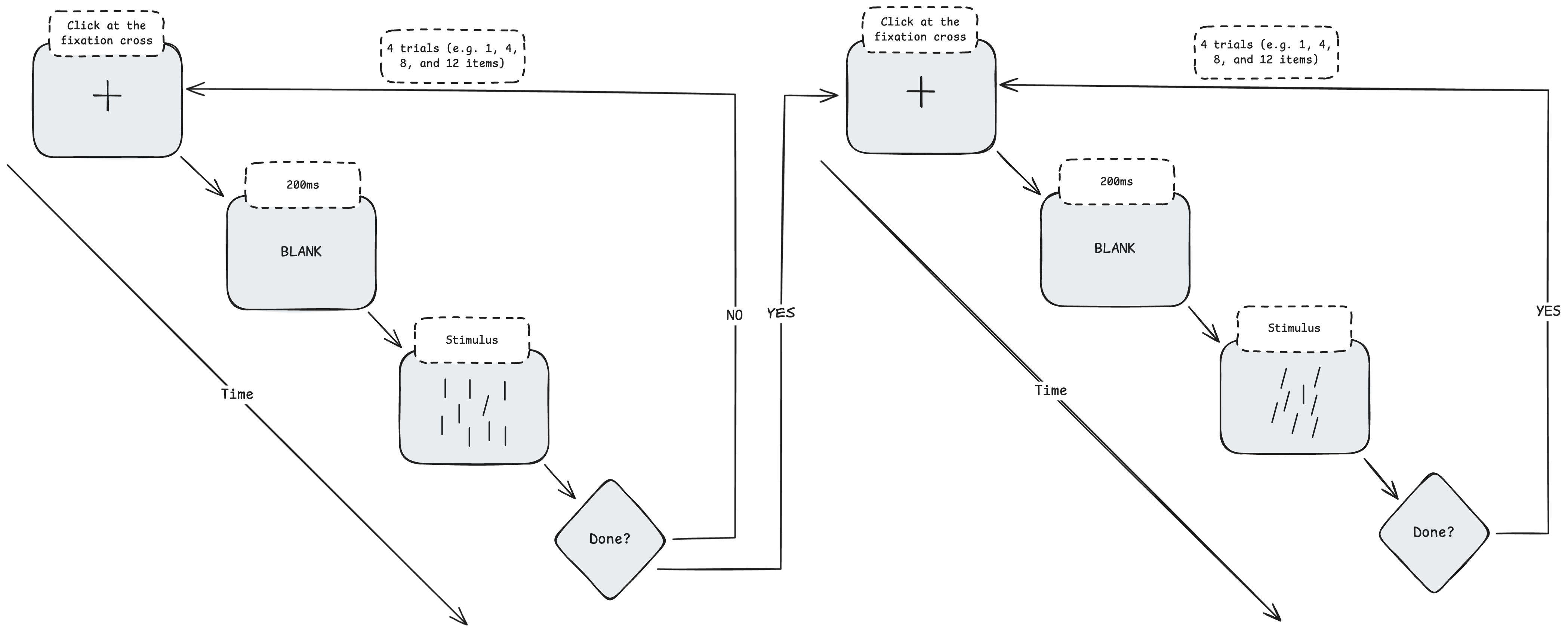}
    \label{fig-exp-sketch}
\end{figure} 
\end{indented}
\subsection{Analysis}
\subsubsection{Data Analysis}
\begin{figure}[h!]
    \centering
    \caption{Box plot showing the raw dataset, before removal of the outliers - Figure \ref{outliers-1}; and then the resulting dataset - Figure \ref{outliers-2}. The blue dots on the second plot are extreme data points, not to be considered outliers.}
    \begin{subfigure}[b]{0.7\textwidth}
        \centering
        \includegraphics[width=\textwidth]{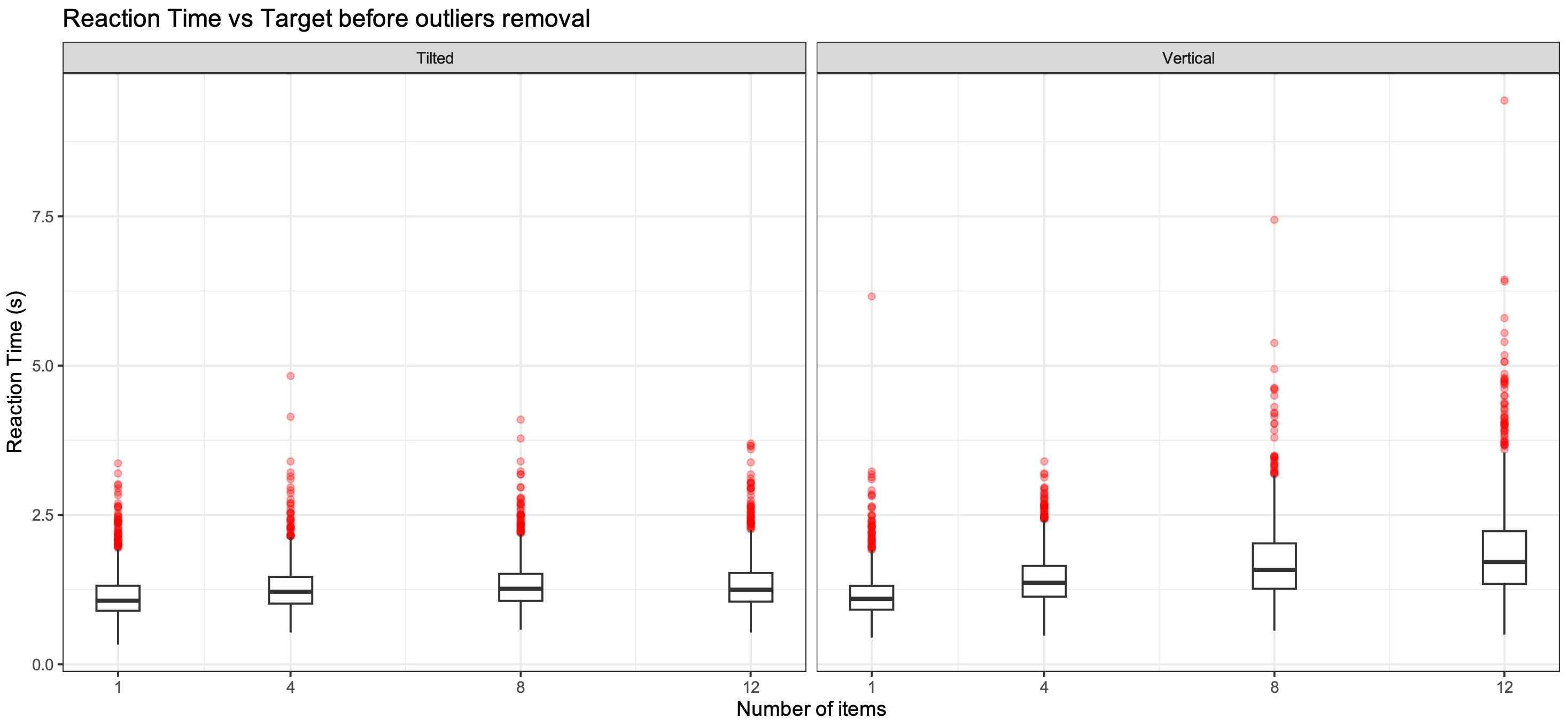}
        \caption{Dataset containing outliers.}
        \label{outliers-1}
    \end{subfigure}
    \hfill
    \begin{subfigure}[b]{0.7\textwidth}
        \centering
        \includegraphics[width=\textwidth]{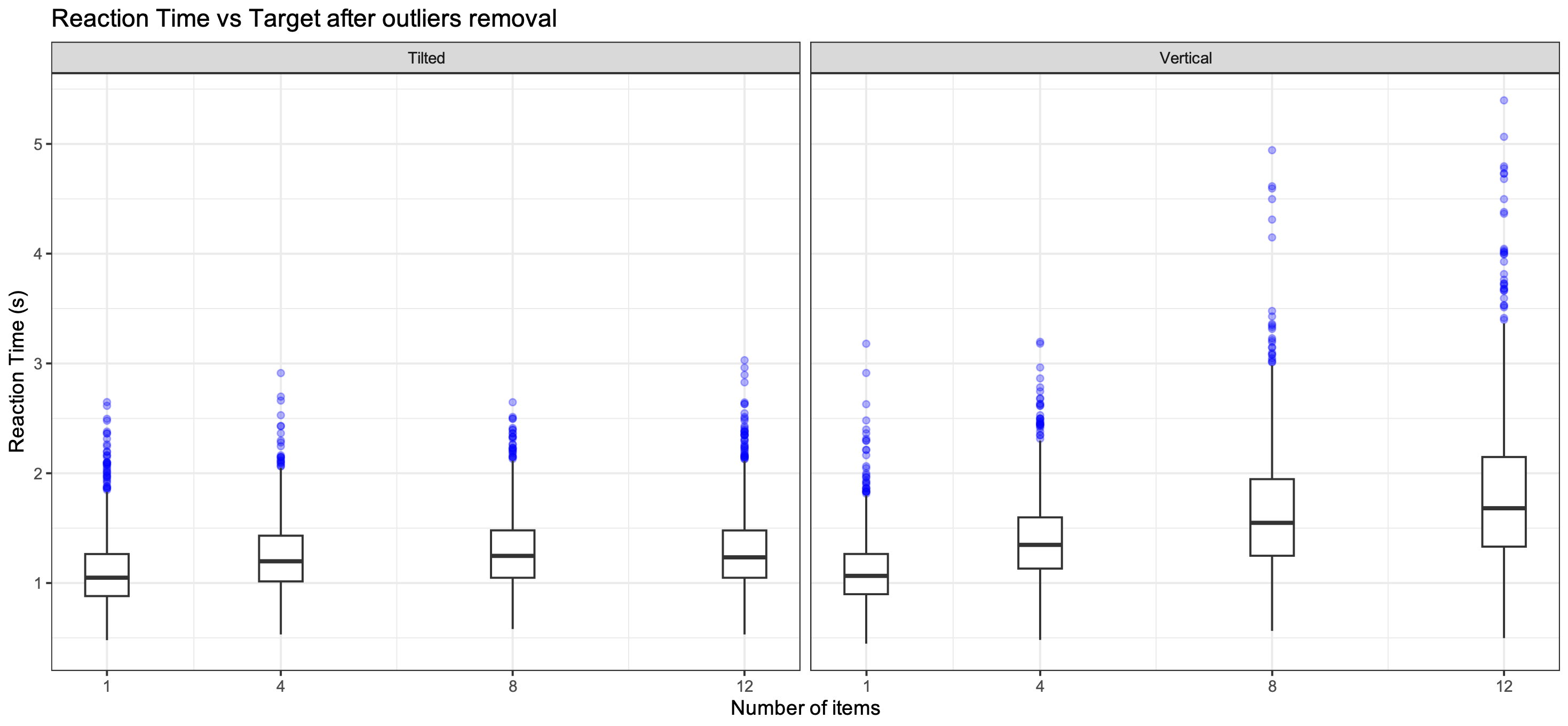}
        \caption{Dataset after outliers removal. Extreme data points are not removed.}
        \label{outliers-2}
    \end{subfigure}
    \label{before-after-outliers}
\end{figure}

Data analysis was performed using the R programming language, version 4.4.1, published by The R Foundation for Statistical Computing on the 14th of June 2024. The \textit{ggplot2} package was used for plotting the charts. We defined a minimum reaction time of 100 milliseconds and filtered out data points below that RT. Other outliers were identified using the Inter-quartile Range (IQR) function, as defined in the Equation \ref{eq-outliers}.
\begin{equation}
    \text{outliers} = x > Q_{\text{3}} + 1.5(Q_{\text{3}} - Q_{\text{1}}) \lor x < Q_{\text{1}} - 1.5(Q_{\text{3}} - Q_{\text{1}})
    \label{eq-outliers}
\end{equation}
\begin{indented}
The IQR function filtered outliers based on participants, number of items, and target type. The raw dataset consisted of 9360 data points; after removing the outliers, the dataset was reduced by 4.91\%, with remaining 8900 data points. Figure \ref{before-after-outliers} shows the data distribution before and after the operation.
\end{indented}
\subsubsection{Statistical Analysis}
We performed statistical analysis to test the hypothesis that oblique targets would be identified faster, while vertical targets will require focused attention, i.e. more time to be identified. The Reaction Time (RT) was the dependent variable and the Target Type (TT) the independent variable. We modelled the RT as a function of TT, as we wanted to identify whether the TT would have a significant impact on RT or not. We performed an ANOVA test, with '$rt \sim tt$' as formula; then another ANOVA test, analysing the interaction between TT and Age, with '$rt \sim tt * age$' as formula; and finally an ANOVA with Bayes Factor, with '$rt \sim tt$' as formula, as we wanted to identify how strong the evidence was.\\
\begin{indented}
Additionally, we applied a Linear Regression Model to study the slope of the RT per number of items and TT. That was used to calculate the intercept of the slopes and the p-value.
\end{indented}
\section{Results}
\begin{figure}[h!]
    \centering
    \caption{Average reaction times of all participants, including the regression lines, across trial blocks with 1, 4, 8, and 12 items.}
    \includegraphics[width=0.8\textwidth]{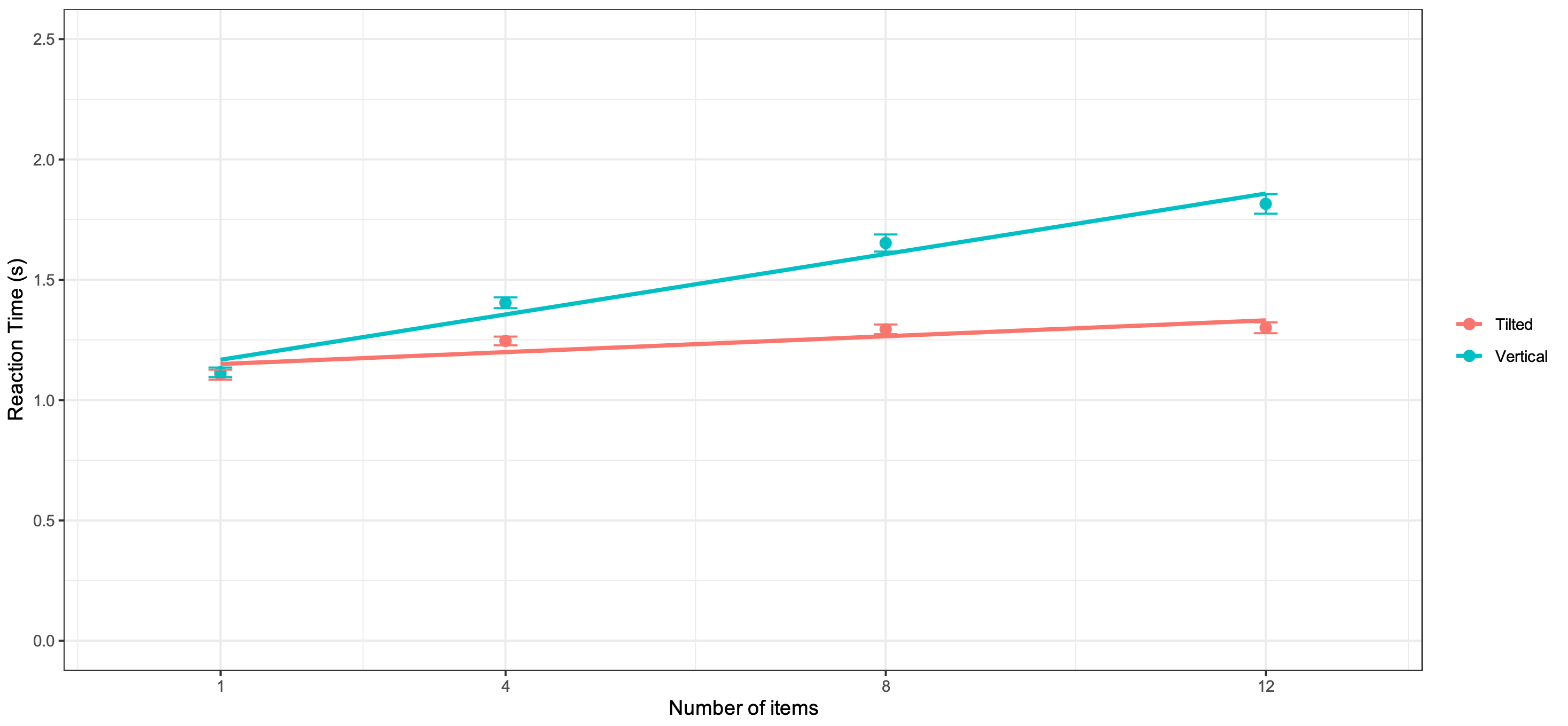}
    \label{fig-sk-results}
\end{figure}

\begin{figure}[h!]
    \centering
    \caption{Regression lines on top of a violin plot and a histogram with Mean and Median vertical lines. Sub-figure (a) highlights the extreme values retained in the dataset despite the application of the IQR function to remove outliers.}
    \begin{subfigure}[b]{0.75\textwidth}
        \centering
        \includegraphics[width=\textwidth]{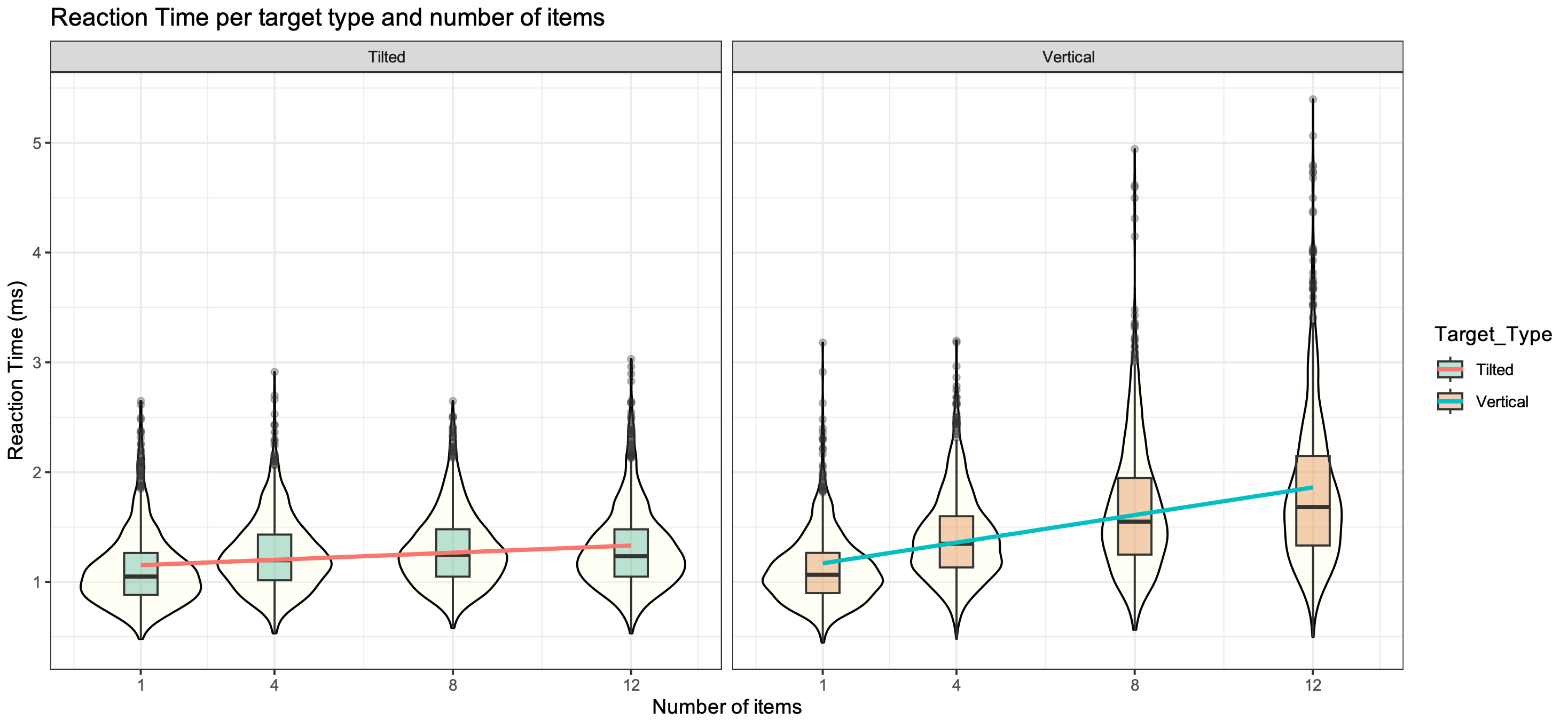}
        \caption{Reaction time is analysed in relation to target type and the number of items, with the data density along the x-axis resembling a violin shape.}
    \end{subfigure}
    \hfill
    \begin{subfigure}[b]{0.75\textwidth}
        \centering
        \includegraphics[width=\textwidth]{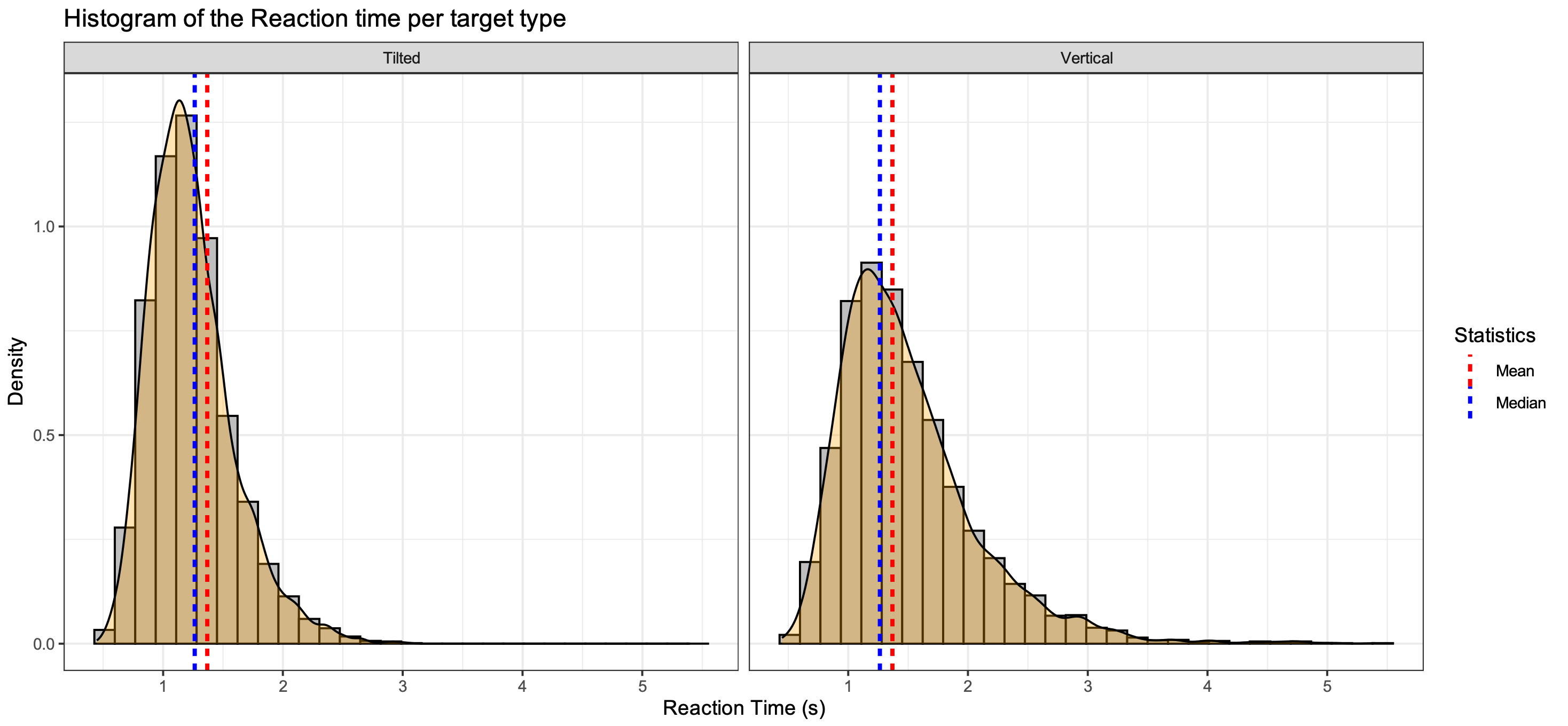}
        \caption{Data density illustrated in a histogram plot, showcasing the Mean and Median RTs for both target types}
    \end{subfigure} 
    \label{violin-hist-plot}
\end{figure}

The ANOVA model outcome was favourable towards our hypothesis and showed that the target orientation had a strong effect on RT. The RT was higher when the target was a vertical line among oblique distractors as opposed to a lower RT with an oblique target. The ANOVA output a \textit{F-value = 666.7} and a \textit{P-value < 0.001}, showing high significance of the TT on the RT at any significance level (e.g., 0.05, 0.01, or 0.001).\\
\begin{indented}
The second ANOVA model, where we took into account the interaction between TT and Age, we observed a high \textit{F-value = 668.525} and a \textit{P-value < 0.0001} for the TT. When looking into the Age model only, we observed a lower F-value, but still a significant effect, with values of \textit{F-value = 17.613} and a \textit{P-value < 0.0001}. The interaction between TT and Age showed a significant effect, but not as high as the isolated models. For instance, we observed a \textit{F-value = 9.135} and a \textit{P-value = 0.00251}. The Figure \ref{fig-sk-results} displays RT slopes across conditions, showing an increase in RT when the target is a vertical line. Table \ref{lm-results} quantifies these results, highlighting a significant effect of target orientation on RT. We further inspected the data distribution with a regression line, as illustrated in Figure \ref{violin-hist-plot}.\\
To conclude our results, the Bayes Factor (BF) analysis showed overwhelming evidence that the TT has a very high effect on the reaction time of the participants. The observed value was $BF = 6.16 * 10^{137}$. We based this results on the work by Kass and Raftery (1995)
\cite{kass}.
\end{indented}
\begin{table}[h!]
\centering
\caption{Linear regression model results. The intercept represents the mean reaction time for the oblique target, with a value of 1.24s. The vertical target shows an increase of 0.26s in reaction time compared to the oblique target.}
\begin{tabularx}{\textwidth}{|>{\centering\arraybackslash}X|>{\centering\arraybackslash}X|>{\centering\arraybackslash}X|>{\centering\arraybackslash}X|>{\centering\arraybackslash}X|}
\hline
\textbf{Target} & \textbf{Estimate (s)} & \textbf{Standard Error} & \textbf{t-value} & \textbf{p-value} \\
\hline
Intercept & 1.237945 & 0.007177 & 172.49 & <0.0001 \\ \hline
Vertical & 0.261824 & 0.010140 & 25.82 & <0.0001 \\ \hline
\end{tabularx}
\label{lm-results}
\end{table}
\section{Discussion}
In this study, we investigated parallel and serial search as phenomena within the pre-attentive 'pop-out' and focused attention 'glue' phases of visual search tasks, following Treisman’s framework (Treisman \& Gelade, 1980) \cite{treisman1}. The 'pop-out' effect, aligned with parallel search, occurs when a target stands out due to basic features like colour, shape, or orientation, while the 'glue' phase involves focused attention as multiple features are integrated to identify the target.\\
\begin{indented}
We conducted trials with varying item numbers under two conditions, replicating Gupta et al. experiment \cite{gupta}. Similar to Treisman and Gupta’s findings, we observed search asymmetry: participants identified oblique targets among vertical distractors faster, with a significant difference in reaction time. This led us to reject the null hypothesis, supporting the connection between 'pop-out' and parallel search.\\
In conclusion, we show evidence to support our hypothesis regarding the use of 'pop-out' or parallel search for oblique targets, consistent with Gupta's experiment. We would expect that with a larger sample size, oblique targets would likely be represented by an almost flat line. Further research involving more complex objects with additional features could provide deeper insights into how we perceive visual elements and where certain phenomena occur within the visual cortex, such as the detection of edges, curves, and whole objects.\\
\end{indented}

\subsection{Conclusion}
To conclude, we are pleased to have gathered sufficient evidence to support our hypothesis, with results comparable to previous studies, such as those by Treisman and Gupta.\\
\begin{indented}
Further research exploring Feature Integration Theory and the similarities between neural networks and the mammalian visual cortex could offer a better understanding of how complex features are perceived. A more effective approach, such as eye-tracking, could help to understand the reaction time to identify features and decrease the number of outliers.
\end{indented}
\bibliographystyle{abbrv}
\bibliography{pop_out_vs_glue}

\end{document}